# Induced Homomorphism: Kirchhoff's Law in Photonics

Shuai Sun[1], Mario Miscuglio[1], Xiaoxuan Ma[1], Zhizhen Ma[1], Chen Shen[1], Engin Kayraklioglu[1], Jeffery Anderson[1], Tarek El Ghazawi[1], Volker J. Sorger*

[1] Department of Electrical and Computer Engineering, The George Washington University, Washington, DC 20052, USA

\* Corresponding author *sorger@gwu.edu*
† These authors contributed equally

**Abstract:**
**When solving, modelling or reasoning about complex problems, it is usually convenient to use the knowledge of a parallel physical system for representing it. This is the case of lumped-circuit abstraction, which can be used for representing mechanical and acoustic systems, thermal and heat-diffusion problems and in general partial differential equations. Integrated photonic platforms hold the prospect to perform signal processing and analog computing inherently, by mapping into hardware specific operations which relies on the wave-nature of their signals, without trusting on logic gates and digital states like electronics. Although, the distributed nature of photonic platforms leads to the absence of an equivalent approximation to Kirchhoff's law, the main principle used for representing physical systems using circuits. Here we argue that in absence of a straightforward parallelism and homomorphism can be induced. Here, we introduce a photonic platform capable of mimicking Kirchhoff's law in photonics and used as node of a finite difference mesh for solving partial differential equation using monochromatic light in the telecommunication wavelength. We experimentally demonstrate generating in one-shot discrete solution of a Laplace partial differential equation, with an accuracy above 95% with respect to commercial solvers, for an arbitrary set of boundary conditions. Our photonic engine can provide a route to achieve chip-scale, fast (10s of ps), and integrable reprogrammable accelerators for the next generation hybrid high performance computing.**

**Introduction:**

Photonic integrated circuits (PICs) do not exist. Even if this statement could seem outrageously contradictory, we invite the reader to bear with us while we unravel the assertion.
The concept of a circuit originates from connecting electronic components into a functional unit and as such is governed by certain physical rules. These fundamental rules of circuitry that, in fact, does not exist in optics and hence also not in photonic platforms. As such, the perception of a circuit applied to photonics is actually only a rather loose sense, with significant physical and technological consequences.

In electronics, circuits are simply loops, in which the flow of electrons circulates, ruled by the conservation laws, governed by quasi-static approximation of Maxwell's equation, resulting in Kirchhoff's law. In contrast, in photonics, light does not have return loops but is usually conveyed to be ultimately detected. Conservation laws still hold considering the light dissipation and transition into other domains, but the 'flow of photons' follows Maxwell's equations and only its transversal model approximation can be applied.

Another main aspect to consider, as a consequence of the quasi-static assumption, is that we usually refer to circuits when we can approximate their components as concentrated at singular points in space ('lumped circuits') in which the physical quantities, such as potentials and currents, are function of time only. This approximation is possible because the wavelength of the signals, and their time-scale variation, is significantly longer and slower than the physical dimension and variations of the circuit itself, respectively. The consequence of this approximation is non-local effects, i.e. elements of the circuits are coupled, and local variations will affect the global performance of the circuit.

In photonics, that is not the case (**Fig. 1 i-iii**); photonic platforms have to be considered as distributed networks since they are characterized by a footprint that is several order of magnitude lager compared to their operating wavelength (Length$_{PIC}$ >> $\lambda_{NIR}$). Typically, when designing 'photonic circuits', the aim is to design components that will generate photons and efficiently convey them, modulated in some prearranged way, to obtain a certain functionality. Thus, photonic circuits are information pathways rather than circuits (**Fig. 1.iii**). For this reason, photonics is considerably more related to data transport than to algorithmic transformations and operations.

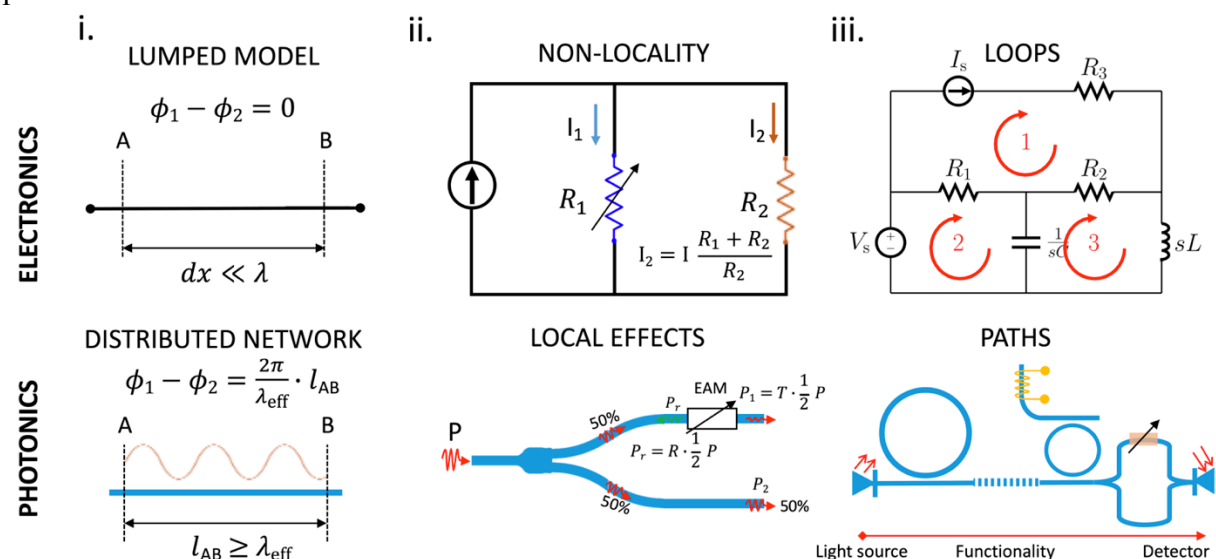

**Figure 1 Fundamental differences between electronics and photonics impacting design and functionalities. (i)** Electric circuits have a footprint significantly smaller than the operating wavelength, thus can be approximated via a lumped model; while in photonics the wavelength of the electric field is significantly smaller than the network. **(ii)** In electric circuits non-locality holds true, where local variations do affect the entire network functionality, while in photonics light intensity is only affected locally (e.g. after being modulated). Output power $P_1$ and $P_2$ of the splitter is given in terms of *T* (Transmittance), *R* is the (Reflectance). **(iii)** Electric current flows in loops, while light is generated, then follows set pathways, which provides a certain functionality, and is ultimately detected.

Therefore, strictly speaking, the "photonic circuit", which confines and manipulates light, defectively performs its functionality on a component-by-component manner. However, Kirchhoff's and non-local effects contribute in mapping in integral way fundamental mechanism used for describing important phenomena directly into electrical circuits and their algorithmic functionality. Indeed, such 'coupled-nodes' as part of a phase-constant network can be a powerful tool for mimicking integral-differential problems, for instance, which ubiquitously pertains to a plethora of diverse scientific and engineering problems, onto photonic hardware, thus potentially relevant for future analog computing accelerators.

To further explain this concept, at the dawn of the digital era, analog processors based on electrical mesh and Kirchhoff's law have been conceptualized and demonstrated(*1–3*). Such analog processors were able to solve second order partial differential equations (PDEs) using finite difference methods (FDM), relying on continuous signals and programmed by changing the interactions between its computing elements, e.g. impedances, using minimum stored programs or algorithms; thus obtaining solutions in a completely asynchronous manner, providing one-time, (non-iterative) computations independent, at first order, of the problem complexity. However, the complexities of an effective integration of a high-speed programmable and energy-efficient static-like analog mesh and the concurrent advancement of digital electronics architectures, eclipse this Kirchhoff's electrical FDM approach.

Current (von Neumann) processors solve PDEs through numerical methods, involving iterative vector-matrix high-bit precision operations, which can be both power and time-costly according to the complexity and resolution of the problem. These bottlenecks are only softened by parallel hardware (i.e. multi-core processing), which do not offer a significantly different path to accelerate PDEs, due to the parallelism overhead and disadvantageous computation complexity scaling (*4*).

Since the late 2000s, the computing paradigm has shifted again; it seems apparent that a new class of hybrid hardware is emerging, i.e. co-processor and accelerator, able to perform a task efficiently, by homomorphically mapping a specific problem category on application-specific hardware, which solves the problem at hand in an entirely parallel manner. Such outlook can potentially ameliorate the computing pressure on digital electronics.

In this regards, integrated photonics-based signal processing, thanks to the electromagnetic nature of its signals and availing their efficient interactions with matter, places itself as an compelling solution for optical communication(*5*), quantum information processing(*6*), computing(*7*) and especially neuromorphic computing with remarkably reduced energy consumption and accelerate intelligence prediction tasks(*8–11*). Recently, inverse-designed metamaterial platform interfaced with integrated photonics showed the possibility of solving integral equations using monochromatic electromagnetic radiation(*12*). In the field of PDE solvers, all-optical reconfigurable module based on micro-ring resonators can solve ordinary first and second-order temporal differential-equation(*13*, *14*).

Here we argue and demonstrate that, to a certain extent, we can induce circuit-related homeomorphism by forcing an integrated photonics platform to behave as a circuit in which Kirchhoff's law can be re-implemented. We experimentally demonstrate a photonic node, termed Kirchhoff's Photonic Node (KPN), which splits the incoming light evenly into three remaining directions with minimized reflections, which can be reprogrammed emulating an optical version of Kirchhoff's current law using optical light intensity. When multiple nodes are interconnected and arranged to generate a two-dimensional mesh, in analogy to a uniform electronic resistive circuit, we are able to approximately solve a stationary partial differential equation which simulates in a non-iterative fashion the heat-transfer problem of a film, via an optical finite difference method (FDM). We observe that when the proposed photonic circuit, here termed Silicon Photonic Approximate Computing Engine (SPACE), is forced to pseudo-homomorphically map the PDE, we can achieve solution accuracy up to 97% compared to a simulated heat transfer problem with the same mesh resolution which can be obtained in just 16 ps.

**Results:**

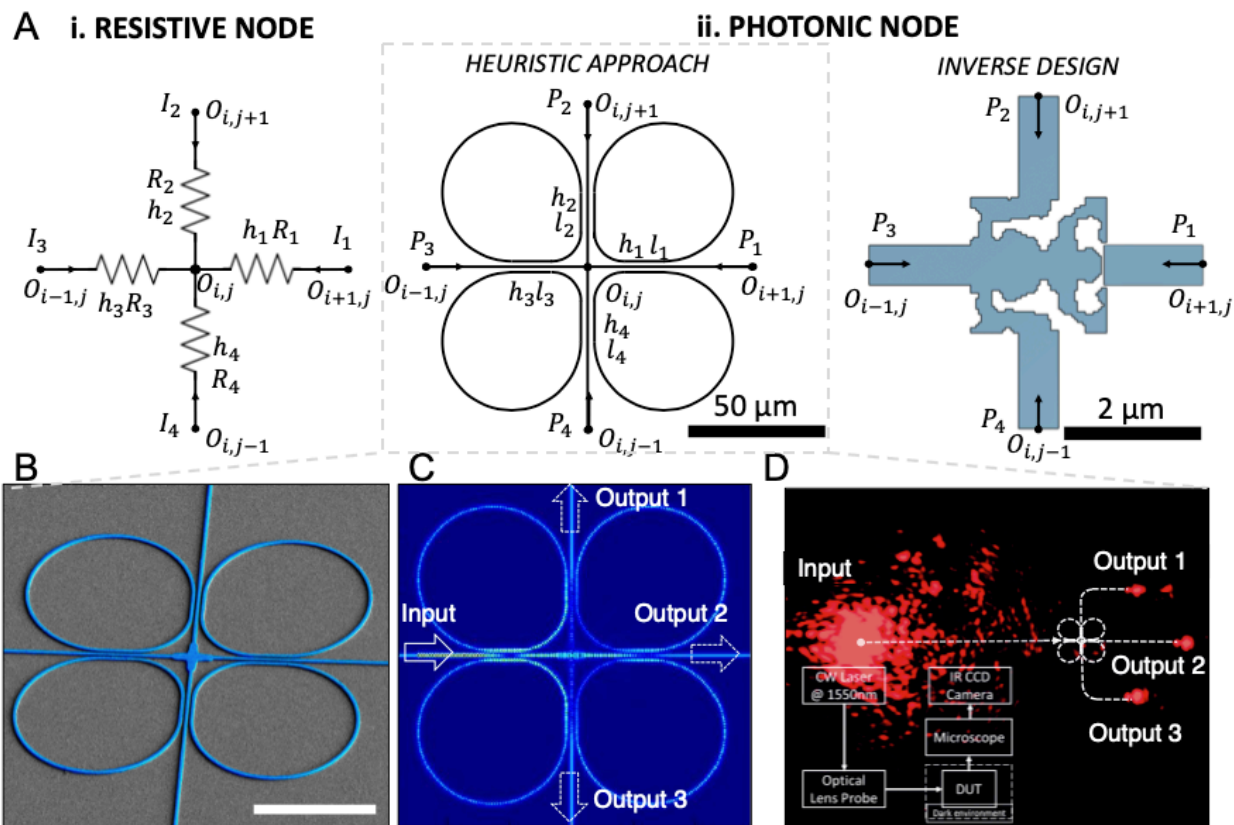

**Figure 2 Kirchhoff's Photonic Node (KPN).** (**A**) Node mapping between (**i**) a resistive mesh and (**ii**) a photonic node, designed either using *heuristic* approach or *inverse design* algorithm. Kirchhoff's Photonic Node implemented either as waveguide crossing assisted by a four waterdrop-like ring resonators or using optimized inverse design(*15–17*) (represented with regions of silicon indicated by blue, and silicon dioxide indicated by white). (**B**) SEM micrograph (53° tilted) of a fabricated Kirchhoff's Photonic Node (KPN). False color highlighted in light blue the Silicon KPN. Scale bar 20μm (**C**) Electric field distribution using 3D FDTD simulation of the KPN. A TM mode source is injected from the left and aided by the crafted coupling coefficients split evenly in the other 3 cardinal directions. (**D**) The microscope image captured by infrared camera assessing the equal splitting with an inset showing our measurement setup. All simulation and measurement use light source at 1550 nm wavelength.

The fundamental unit of an electrical circuits are the nodes, which represent the terminals in which two or more circuit elements meet, and their distribution and interconnection determine circuit functionalities and operating conditions. At the node of electrical circuit, Kirchhoff's current law applies; which states that current flowing into a node (or a junction) must be equal to current flowing out of it. This is a consequence of charge conservation (energy), if we can concurrently assume that the instantaneous variation of the magnetic flux outside a conductor and the change of charge in the conductor is zero (steady-state conditions). In photonics, due to the distributed nature ($l \geq \lambda_{\text{eff}}$) of the platform used, the behavior of the electromagnetic radiation is governed by Maxwell's equation in time-variant conditions, in which the electric and magnetic field are both function of time and position (**Fig. 1**). Hence, for our analysis we consider as fundamental quantity of the photonic circuit the optical power instead of electromagnetic field intensity. Although, this is not sufficient for ensuring that the electromagnetic radiation flowing into a node is necessarily equal to the one coming out it. This can be primarily due to 1) impedance mismatch at the discontinuity (i.e. joint between different network elements), which causes reflections, thus producing interference with the incoming radiation, 2) light scattering at the abrupt discontinuities, and 3) optical losses due to mode dispersion. Therefore, prototyping a photonic node in which unwanted reflections and optical losses are minimized would guarantee a photonic equivalent Kirchhoff's law applied to optical power. As a first step towards this goal, we aim to mimic an electric node (**Fig.2 A.i**) with equal resistors onto a photonic platform, which evenly partitions the optical power flowing from one of its side.

To achieve equal light splitting, this node design needs to meet the following three criteria: a) to be symmetrical to both x- and y-axis in order to physically build the scalable optical mesh, and needs to provide a 1-to-3 equal splitting ratio; b) the splitter needs to have tolerance to the fabrication variance since cascading the node will amplify the device variance; c) the segment for light coupling should have the potential to be further integrated with tuning mechanisms (e.g. electro-optic means) in order to ensure reconfigurability and compute-programability. Independently of the photonic node typology, the optical loss along each light path can be used as an equivalent resistances $R_i$ in the electrical model.

For this aim, we used two distinctive approaches; Firstly, we follow a heuristic approach in which to obtain even splitting of the optical power using waveguide crossing assisted by four directional couplers integrated with ring resonators (**Fig. 2 A.ii**). To achieve the design of the KPN we use a heuristic process to obtain 1-3 equal power splitter in photonics by iteratively optimizing the splitting ratio using 3D full-wave numerical simulations. The resulting design comprises of four water-drop shaped rings placed close to two perpendicularly crossed waveguides to couple part of the light coming from one direction into both, the other two perpendicular directions and still let the remaining light pass through to the opposite port. Instead of using circular rings, the segments close to the straight waveguides are flattened to form a three-waveguide directional coupler. We used directional couplers to couple into the 4 feedback loops, and refrain from using neither perfectly circular rings and nor high-quality factor cavities to widen the spectral (and thermal) operating window such as to not having to use tuning (e.g. thermal, electro-optic) to control its resonance. In addition, a 4-way waveguide crossing is the center of each Photonic Kirchhoff's node to reduce the scattering and crosstalk at the intersection.

Secondly, an optimized inverse design approach is used by setting the design area to 5 μm and the even splitting functionality in the cardinal directions according to the following cost function to optimize

$$\min T_{\text{obj}} = |T - Pout_2| + |T - Pout_3|^2 + |T - Pout_4|^2, \qquad (1)$$

where T is the target transmittance and $Pout_n$ is the power at the port *n,* assuming the following constraints:

$$0 \leq Pout_2 + Pout_3 + Pout_4 \leq 1 \; (Pout_1) \qquad (2)$$

In the full space of fabricable devices, the optimization algorithm finds a structure (**Figure 2A.ii**) that meets these requirements. (Further details of the electromagnetic characterization of the inverse design KPN can be found in the SOM, **Fig. S2**).

The inversely designed KPN is one among the infinite number of configurations that would satisfy the optimization of the cost function, and even if characterized by a compact size (2x2 μm), it is significantly more intricate to fabricate with the same high yield as regular photonics due to its limited size. Additionally, this type of node would require a completely different configuration and related optimization process for mimicking different 'optical power' partitioning, while the heuristic solution can be straightforwardly reconfigured by actively tuning the coupling coefficients with the ring resonators. It is worth noticing that the inverse design of a KPN is not symmetric, therefore the S-parameter matrix is not equal to its transpose ($S \neq S^T$) leading to a non-reciprocal behavior, meaning that the inverse designed KPN would produce different splitting ratios and reflections according to which port the source is assigned. For these reasons, we decide to continue our analysis, device fabrication and testing using the KPN obtained through heuristic approach.

After optimizing the bending radii of the water-drop shaped rings, flattened coupler length and the gap between the ring and the straight waveguide, the splitting ratio can be tuned to 22%, 23% and 22% with 12% reflection (**Fig. 2 B**) based on full-wave simulation (Lumerical 3D FDTD). Here the reflection is mainly caused by the return couplings from the three-waveguide couplers at the perpendicular ports (i.e. Output port 1 and 3). Instead of completely coupling to the perpendicular port, the light coming from the first two rings will be partially leaked to the rings on the other side and route the signal back to the input port. In terms of the fabrication process, 2% hydrogen silsesquioxane (HSQ) electron-beam resist is used due to its fine resolution and edge contrast, with isotropic dry etching process ($SF_6$ and $C_4F_8$) to get the uniform height profile and vertical sidewall profile. More details related to the fabrication are given in the Method section.

In order to test the fabricated device, a 1550 nm continuous wave (CW) laser is used as light source which is coupled to the waveguide by means of a periodical grating coupler (Details regarding the grating coupler are discussed in the Supplementary Online Material). To read out the output values at each port of the KPN in parallel, an InGaAs infrared camera (Xenics) is used to capture the microscope image of the light outcoupled in each direction with 14 bits ($2^{14}$ levels) of precision. The background noise, such as the arbitrary reflections from the sample surface and the sensor thermal noise, was minimized using noise-canceling method and post image processing, thence the light intensity from the grating coupler regions in each direction was acquired (More details are provided in the Method section) (**Fig. 2 C**). As the result, the light intensity from all three output ports have a ratio of 22.3% : 23% : 22.1% (or 1744.5 : 1801.6 : 1727.2 from the image pixel readout) which is in excellent agreement with the FDTD simulation result with less than 0.5% deviation. Nevertheless, we envision that high-speed, low noise germanium(*18–20*) or graphene photodetectors(*21–24*) can be integrated into the device and used for improving both detectability and data collection speed and accuracy.

To showcase the functionality of a photonic platform formed by KPNs, we aim to approximate a finite difference node, which locally discretize a Laplace equation using a finite difference method (more details in the **SOM Section 2-3**). Although, this is not the only possible application space for the proposed architecture since a network comprising of KPNs could be used as recurrent neural network(*25*) or as a compact solution for reconfigurable routing and network broadcasting (*26*), or also simply as a reprogrammable filter for information pre-processing applications such as for network-edge devices.

As a proof of principle, here we select a 2-dimensional heat transfer problem represented by a steady-state Laplace's homogeneous equation (**Fig. 3A i**), which can be mathematically described by **Eqn. 1**, which describes the relation between a variable *f* and its partial derivatives. Typically, PDEs are solved numerically by discretizing space (and/or time) into meshes points, in such a way that the partial derivatives can be reduced into linear combinations of the variable values at several neighboring nodes of the mesh.

In details, after applying the Finite Difference Method (FDM) to a mesh network (**Fig. 3A ii**), the central node $O_{i,j}$ can be represented by its four adjacent nodes (**Eqn. 2**), where $h_i$ is the mesh step that describes the discretization level of the problem in the analytical domain. Once the discretized mesh node is set with node-to-node correlation function $\epsilon_i$ approximate to a constant value when the equidistant mesh step $h$ is small enough, this Laplace's equation can be locally converted in summation of incremental ratios of physical quantities and solved iteratively. (Further details in SOM **Section 2**)

However, this usually requires a large amount of compute power, memory, and scales exponentially as the problem size and required accuracy.

$$\nabla^2 f = \frac{\partial^2 f}{\partial x^2} + \frac{\partial^2 f}{\partial y^2} = 0 \tag{3}$$

$$\nabla^2 f \cong \frac{\epsilon_1(f_1-f_0)}{h_1^2} + \frac{\epsilon_2(f_2-f_0)}{h_2^2} + \frac{\epsilon_3(f_3-f_0)}{h_3^2} + \frac{\epsilon_4(f_4-f_0)}{h_4^2} \cong \frac{1}{h^2}(f_1 + f_2 + f_3 + f_4 - 4f_o) \tag{4}$$

As suggested more than 6 decades ago (*1*), electrical resistor networks (**Fig. 3A iii**) can one-to-one map finite difference mesh grids. Similar to the analytical model, electrical current $I_i$ and resistance $R_i$ can be mapped to such a node model ruled by Kirchhoff's law and Ohm's law written as $I_i=(V_i-V_o)/R_i$. Here, the current injected into the selected node always equals to the current leaving the node (Kirchhoff's law), while the current splitting ratios to each direction will be automatically 'adjusted' by the intrinsic electrical potentials of each path according to current (voltage) partitions, thus providing a solution to the problem (e.g. PDE). (More details in SOM **Section 2-3**).

While the mapping in an electrical circuit is completely homomorphic, in photonics due to its distributed nature this is not possible. In electric circuits, due to limited dimension favoring at low operating speed quasi-static approximation, a variation of just one resistance, which represents the node-to-node correlation in the discretized domain, induces a redistribution of the potential in the entire network (non-local effects), which allows to model different gradient effects in the distribution of physical quantities. In photonics instead, the optical power when flowing in the photonic mesh is not affected by perturbation in other paths in a global sense, though this effect can be forced or induced. In order to approximate the potential distribution in different lumped circuits operating at steady state, one can build a look-up table which comprise diverse photonic meshes that approximate their behavior. However, addressing all the splitting ratios at each node could be unpractical under certain circumstances, adjusting only the "key nodes" (e.g. nodes on the boundaries or high-loss nodes) which located adjacent to the places that the node connectivity has relatively high variations would already bring the accuracy to an acceptable range (a boundary-weighted example and the method of generating key node configuration library is given in **Section 6 and 7** of the SOM in more details). In this view, a photonic processor, with induced homomorphism, could generate an initial, low-precision and approximate guess, to be used then by an integrated high-precision digital solver which ultimately produce the required high-precision solutions. This type of photonic engine can, indeed, configure itself as an accelerator as it would reduce the number of iterations required by the digital solver when solving partial differential equations associated to complex numerical simulation in an iterative fashion, e.g. Newton's method. Practically, for this kind of solvers, it is essential to obtain an initial approximation to the solution which can be used as an initialization stage for obtaining a fastly convergent simulation.

Nevertheless, besides the disadvantage of populating the space of configurations to mimic different lumped circuit behaviors, using integrated photonics would provide three immediate advantages over resistive networks. First, the absence of charge/discharge of the wires enables distributed communication and concurrent low power dissipation. Second, once the network is set, the picosecond-short delay is dominated by the time-of-flight of the photon within the PIC. Third, the amplitude (power) of the traveling wave in the waveguides can be easily tuned using attojoule-efficient modulators through wisely engineered light-matter interactions (*27–30*) mimicking different configurations (programmability). For the sake of simplicity, we demonstrate that it is possible to obtain accurate solutions for a uniform domain, but this approach could be extended exploiting the network reconfigurability. In the supporting information, exploiting the symmetry and reciprocity of the KPN, we show that the same chip can be used for obtaining the solution of the same Laplace PDE with different Dirichlet's boundary conditions (SOM **Section 5**, **Fig. S10**).

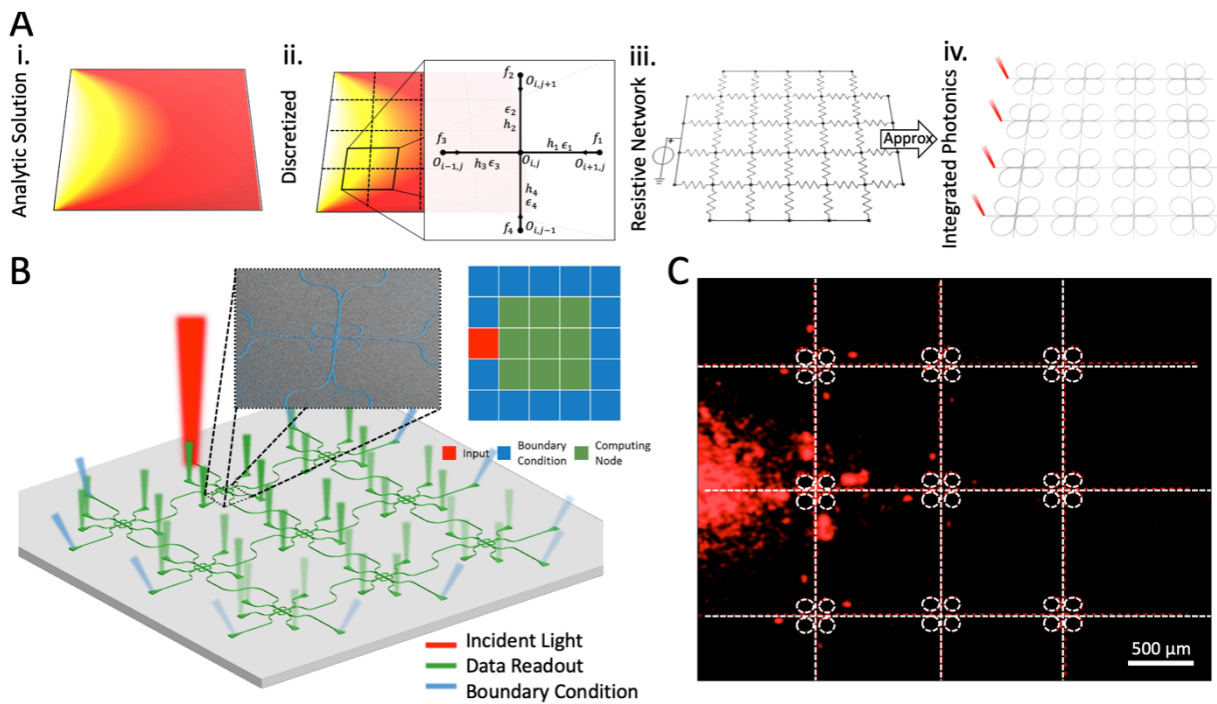

**Figure. 3. Solving PDE using Kirchhoff's Law in integrated photonics**. (**A**) **i.** Analytic solution of a partial differential equation for the defined boundary conditions. **ii**. The discretized solution of the same PDE using numerical methods (finite difference). The overlaid mesh denotes the discretization. Inset highlights the node of a mesh. (Notation is the same as **Fig. 2**) **iii.** Characteristic electrical resistor mesh which maps the finite difference method applied to the PDE. **iv.** A photonic network which imitates the behavior of a lumped circuit obtaining approximate (~97% accuracy, **Fig. 4A**) discretized solutions to the PDE. The discretization step for each solver is considered the same ($n = 3$) and the boundary conditions are applied as external bias voltage or optical power for the electrical and the photonic engines, respectively. (**B**) Schematic 3D demonstration of a heat transfer problem with light injected from the central left. Boundary condition is set by using extended waveguides and grating couplers connect to the peripheral nodes. Light coupled into this direction can be scattered into free-space without reflections thus can be regarded as a perfect constant temperature boundary condition. The block chart shows a top view of the initial setup of the heat transfer problem that can be solved by our 5×5 SPACE design. (**C**) The microscope image captured by infrared camera at 1550 nm wavelength overlaying with a sketch of the optical power splitters. Note, grating couplers, y-branches and bending waveguides are omitted for better visualization.

After providing a practical exhibition and guidelines for obtaining a Kirchhoff's equivalent law in photonics and consequently obtaining an FDM-like node, we cascaded multiple nodes building a 5×5 optical FDM mesh grid to solve a discretized heat transfer problem. The assembled system maps a symmetric type of heat transfer problem with a heat source injected from the center-left of the mesh grid and surrounded by constant temperature boundary**.**

The input signal, which in this case represents the Dirichlet's boundary conditions may, in general, be any arbitrary laser beam distribution coupled into any node of the circuit (i.e. here grating couplers used, **Fig. 3B**). For mimicking a 3x3 FDM mesh a 5x5 photonic mesh is fabricated and tested, in which the additional nodes on the sides of the domain are used for reading/applying the optical power at the boundary. In electrical circuits, a Dirichlet's boundary condition is provided by a constant potential which could be either "active" (heat source), using electrical sources (voltage or current generators), or passive boundary conditions (heat sink) as electrical ground. The electrical node at the boundary has two functions: (1) sets the value of the function at the boundary by applying a constant voltage (2) due to the lumped nature of the circuit, forces a gradient trend (differential voltage) throughout the lumped circuit. In a photonic network, if the former is possible by applying a fixed optical power, the latter is not straightforwardly achievable due to the absence of non-local effects. For these reasons, the gradient is directly embedded into the network by modifying the node-to-node optical losses (i.e. 1 dB of loss in this case by using waveguide bending) mirroring, for the optical power, a linear node-to-node correlation function $\epsilon_i$. This is an additional step towards our quest to an induced homomorphism.

To characterize the performance of the system and obtaining discretized measurements for each node, first, we introduce for each direction of the nodes a set of 50/50 Y-branch splitter followed by a grating coupler in order to estimate the optical power at each node. The power drop at each node represents the temperature distribution at each point of the discretized domain, and it is measured, as previously observed, in time-parallel through a properly calibrated camera (**Fig. 3C**). However, for high-speed reconfiguration operation, integrated phtodetectors can be used with a latency on the order of 10's ps.

In order to obtain readable data from the furthest node from the input, 39 mW of laser input (as the maximum power output from our laser source) is applied to the 5×5 SPACE mesh grid. The optical non-linearity effect has also been considered and the actual optical power coupled into the first node is well below 5 mW to prevent this (*31*).

We verify the accuracy of the approximate solution of the 5×5 SPACE prototype by comparing the obtained experimental measurement at each KPN to the discretized and normalized solution of a heat transfer problem obtained through ac commercially available numerical solver (COMSOL Multi-physics) with the same mesh resolution and density. Considering the different input units (i.e. temperature and optical power) and values (i.e. 100ºC in the thermal model and 39 mW optical power in the measurement), all the output results are normalized in order to represent identical temperature distribution in the discretized domain with same scale while maintaining the same equivalent node-to-node correlation functions $\epsilon_i$.

A discretized solution is obtained through commercially available software (COMSOL) which served as a problem accuracy baseline. It is worth mentioning that any discretized solution is proportional to the mesh resolution (e.g. a 5×5 mesh COMSOL simulation has 99.95% accuracy comparing to a 300×300 mesh averaged down to a 5×5 with same initial setup, **Fig. S3**). In details, in order to decouple the discretization error from the error produced by the approximation of the model, a 5×5 discretized mesh is used as comparison with the experimentally measured solutions at each node of the 5×5 SPACE (**Fig. 4. A**). SPACE provides an overall average accuracy of 98% and a discrete solution with an accuracy higher than 95% for each node (**Fig. 4 B**).

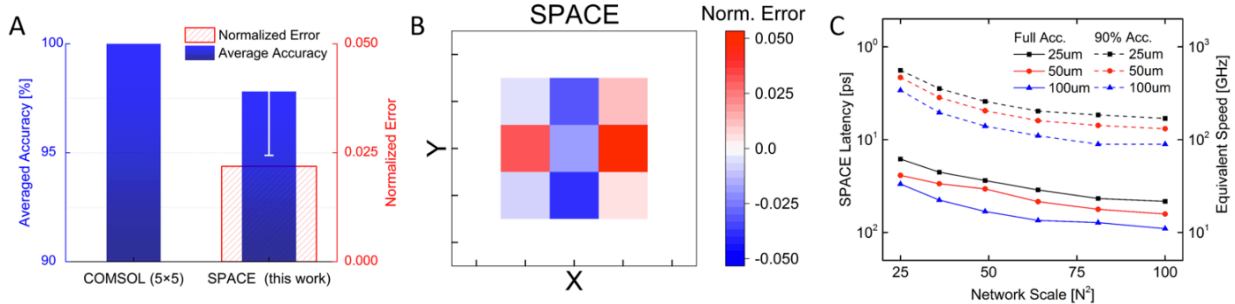

**Figure 4. Accuracy and runtime performance of SPACE when solving Laplace equation mapping a 5×5 FDM.** (**A**) The averaged error and accuracy comparison between COMSOL simulated model (mesh size 5×5) and the measured solutions at each node of a 5×5 SPACE. The negative error bars represent the accuracy level from the least accuracy node from the 5×5 FDM model. The numerical simulation is regarded as the baseline and scaled to 100% accuracy. The accuracy of the solution provided by SPACE is on average 97.5%. (**B**) Normalized error heatmap between the baseline model and the measured space in the scale of (-0.05, 0.05). (More details in the SOM, Section 5) (**C**) Latency analysis for obtaining a stable solution on different network scales from 5×5 to 10×10 with different node-to-node distance varying from 25 μm to 100 μm. Both full accuracy and 90% accuracy runtime show exponential increase in the runtime mainly caused by the node-to-node distance. With closest packing (25 μm), full accuracy and 90% accuracy are able to provide 63 and 556 GHz operating speed respectively. The full accuracy and 90% accuracy are respected to the maximum accuracies that each network scale could get. In all PIC simulations, the input light source has been set to 1 mW with optical power meter sensitivity set to -100 dBm and simulation time long enough to converge all the signal propagation delays in the network.

In our pursuit of a forced homomorphism, similarly to the tolerance of the resistors in a resistive network that mimics a FDM (*1*), the deviations caused by fabrication variances, such as the grating coupler efficiency, y-branch splitting ratio, and variability of coupling coefficients perceptively lower the accuracy of the solution.

Due to the distributed nature of the photonic engine, several iterations, consisting of multiple reflections at each discontinuity, are needed in SPACE to obtain a time-stable solution. This time is proportional to the size of the network and the density, i.e. number of KPN. Based on PIC simulations performed using Lumerical Interconnect, a full iteration cycle (dominated by the time of flight of the photon) takes only 30 ps considering 100 μm node-to-node (n2n) spacing for a 5×5 SPACE in this sparse design. The iteration time drops to 16 ps with 25 μm n2n spacing which is the highest density achievable in our current design (**Fig. 4C**). The highest density can be achieved by using on-chip integrated photodetectors as a detection mechanism, instead of a IR-camera which requires outcoupled radiation from the chip by means of taps and grating couplers at each node which compel sufficient space allocated on chip. Furthermore, considering the photonic node size, the SPACE engine can be packed with a minimum density of 25 μm/component, although we separate the nodes with 200 μm spacing for reducing the output crosstalk while measuring. The footprint of the network can be further shrunk using inverse design approaches obtaining a density of < 5μm/node, enabling higher density meshes. Additionally, SPACE can be fabricated with an adaptive mesh, with an increased density within certain sensitive or turbulent regions of the simulation, thus increasing the overall accuracy. It is also worth to mention that, as an approximate computing engine, when the target accuracy is relaxed to 90% of its maximum, the iteration time drops to 1.8 ps which is equivalent to 556 GHz. In terms of the scalability, the latency saturates as the network size scales from 5×5 to 10×10, in which the light propagation time in the waveguide will contribute even less to the total runtime, thus proving that SPACE could be potentially further scaled-up. (Further details in **SOM**, **Section 5**)

Our KPN design and SPACE circuit provide a powerful tool for homogenously distributing optical power in a defined network similarly to a lumped circuit subjected to Kirchhoff's Law. When the correlation function between each node of the network is wisely selected or actively tuned (e.g. with electro-optic modulators or switches), the network, mimicking FDM, can solve a general second-order Laplace's PDE in an analog manner. For instance, if the splitting ratio of nodes adjacent to the boundary conditions could be tuned to 10% : 10% : 80% (i.e. more light routes to the boundaries), a 5×5 modulated SPACE is able to improve the accuracy up to 99.2% (more details discussed in **SOM, Section 6**). Similar configurations may potentially be explored to solve time dependent or nonlinear PDE with arbitrary boundary conditions by introducing time discretization or nonlinear elements, respectively. PDEs applied to non-homogeneous domains can be mapped on SPACE by changing the splitting ratio according to characteristic distribution (e.g. different thermal conductivity mapped as different attenuation for each individual node). Other cases of PDE applied to non-symmetrical or inhomogeneous domains are reported as PIC numerical simulation in the Supplementary Online Material.

Towards dynamic problem reconfigurability, if electro-absorption modulators(32-34) are introduced between neighboring nodes in SPACE, a vast number of different PDEs can be solved. For illustrative purposes, we numerically show, using a photonic interconnection emulator, that it is possible to map the temperature distribution, solution of Laplace equation, onto SPACE by adjusting the extinction ratio of these modulators between neighboring nodes according to the problem to be solved. This allows solving a multitude of problem cases for number of exemplary underlying heat-conducting materials (Details in the SOM, **Section S7**). Similar to commercial numerical solvers, we present how KPN and SPACE with added reconfigurability produces look-up table solutions for the specific configuration

Beyond the exemplary Laplace equation investigated thus far, other PDEs can, in principle, be solved such as Poisson's equation, for example, if additional light sources are added to the nodes, mimicking the different node potential(*35*). Nonetheless, other PDEs like diffusion equations and wave equations would require optical capacitive and inductive elements needed to express the time-dependent variances enabling the one-shot solution. Different from emulating an optical resistor, which can be easily realized by optical lossy materials or electro-optical modulators, optical capacitors and inductors require specific designs to mimic the behavior of their electrical counterpart. For example, a Fabry-Perot interferometer with chirped Bragg gratings have been demonstrated as an optical capacitive component which can act like a broadband low pass or high pass filter(*36*). On the other hand, an optical inductive component can be implemented as a self-electro-optical device with both integrated modulator and detectors that use the photocurrent to back feed the modulator and change the light intensity injected into the detection region as a negative feedback loop(*37, 38*).

**Conclusions**

In summary, we propose the designs of a photonic node which is able to replicate the equivalent of Kirchhoff's law for optical power. Using equal splitting functionality, we replicate a mesh structure which approximate a homogeneous lumped circuit model. We use the photonic circuit to map a finite difference approach to solve partial differential equation effortlessly and noniteratively, termed Silicon Photonic Approximate Computing Engine (SPACE). Our numerical and experimental analysis indicates that the steady-state response of a characteristic SPACE engine may be achievable in 16 ps, obtaining inherently discretized solutions for each point of the mesh of the domain with a bandwidth up to 63 GHz. This approach could easily adapt another active component, such as electro-optic modulators, photodetectors, and tunable photonic cavities like photonic crystals, to solve more complicated problems with denser, heterogeneous meshes and arbitrary boundary conditions.

Furthermore, the proposed approach features reconfigurability of the input positions and boundary conditions with low-loss network interconnectivity of such distributed networks via PICs and ensures foundry-near cost scaling. This approach allows solving more intricate problems such as those with heterogeneous grid and Neumann boundary conditions when chip is augmented by active components, such as electro-optic modulators, photodetectors, and tunable photonic cavities like photonic crystals. Our findings provided a novel pathway to ultrafast, integrable, and reconfigurable photonic analog computing engine based on an integrated photonics Kirchhoff's node, used for solving PDEs, but can also be adopted as core structure in recurrent neural networks or as compact solution for network broadcasting.

**Methods**

**Fabrication Process** All of the single optical power splitters and 5×5 SPACE is fabricated on the same 220 nm Silicon on Insulator (SOI) chip to minimize the variance during the fabrication process. Raith Voyager 50kV E-beam lithography system is used with fix beam moving stage (FBMS) feature to allow zero waveguide stitching errors across multiple write fields. Hydrogen silsesquioxane (HSQ) with 2% concentration is used to provide around 42 nm of mask thickness (4000 rpm for 60 seconds) with high resolution in writing. After the spin coating, the chip is put on a hotplate for 240 seconds pre-bake at 80-degree centigrade. After the patterning, the chip is dipped into MF-319 for 70 seconds to develop the unexposed HSQ area including 5 seconds of gentle stirring to shake off the air bubbles of the chemical reaction. Then 30 seconds of D.I. water rinse will be immediately applied to stop the development and clean up the residue. To etch down the silicon layer and reveal the features, a 28 seconds of $SF_6$ and $C_4F_8$ (both at 10 sccm) at 500 W ICP power and 20 W bias etching with Plasma-Therm Apex SLR Inductively Coupled Plasma Etcher is able to fully etch all the silicon down and provide over 9:1 selectivity for our smallest features.

**Measurement and Data Processing** To measure the output light intensity, an optical probe station setup is used with a tunable laser at 1550 nm wavelength connecting to a lens fiber to maximize the light coupled onto the chip. Considering the polarization of the grating coupler and its coupling efficiency, the actual laser power coupled into the mesh is less than 5 mW, which is still far below the nonlinearity energy density limitation of the Silicon Photonic waveguide (500nm × 220nm). Xenics IR camera integrated with the microscope captures the scattering light at each output grating. In addition, a black light shield is applied to cover the entire camera, probe station and microscope to prevent the ambient light. And the thermal noise of the camera is eliminated by capturing the image with no laser input. The last type of noise taken into account in the measurement is the surface reflection including the lens flare, and this is by substituting the averaged background readout that adjacent to the grating coupler. After the noise cancelation, the images are imported into Matlab to integrate the intensity values (0〜4095 for our 12-bit depth sensor) of all the pixels of the output region. It is also worth to mention that nodes at different positions have over 3 orders of magnitude difference which is far beyond the dynamic range of the camera. Therefore, lower input laser power with shorter camera integration time is used for nodes closer to the input node and post-processed into the same scale.

**References and Notes:**

1. G. Liebmann, Solution of Partial Differential Equations with a Resistance Network Analogue. *Br. J. Appl. Phys.* **1**, 92–103 (1950).

2. D. C. Barker, Electrical analogue for a partial differential equation. *Mathematics and Computers in Simulation*. **16**, 38–45 (1974).

3. B. S. Vimoke, *Simulating Water Flow in Soil with an Electrical Resistance Network* (Agricultural Research Service, U.S. Department of Agriculture, 1962).

4. G. M. Amdahl, in *Proceedings of the April 18-20, 1967, spring joint computer conference* (Association for Computing Machinery, Atlantic City, New Jersey, 1967; https://doi.org/10.1145/1465482.1465560), *AFIPS '67 (Spring)*, pp. 483–485.

5. Towards systems-on-a-chip. *Nature Photonics*. **12**, 311–311 (2018).

6. X. Qiang, X. Zhou, J. Wang, C. M. Wilkes, T. Loke, S. O'Gara, L. Kling, G. D. Marshall, R. Santagati, T. C. Ralph, J. B. Wang, J. L. O'Brien, M. G. Thompson, J. C. F. Matthews, Large-scale silicon quantum photonics implementing arbitrary two-qubit processing. *Nature Photonics*. **12**, 534–539 (2018).

7. D. A. B. Miller, The role of optics in computing. *Nature Photonics*. **4**, 406–406 (2010).

8. M. Miscuglio, A. Mehrabian, Z. Hu, S. I. Azzam, J. George, A. V. Kildishev, M. Pelton, V. J. Sorger, All-optical nonlinear activation function for photonic neural networks [Invited]. *Opt. Mater. Express, OME*. **8**, 3851–3863 (2018).

9. R. Amin, J. K. George, S. Sun, T. Ferreira de Lima, A. N. Tait, J. B. Khurgin, M. Miscuglio, B. J. Shastri, P. R. Prucnal, T. El-Ghazawi, V. J. Sorger, ITO-based electro-absorption modulator for photonic neural activation function. *APL Materials*. **7**, 081112 (2019).

10. Y. Shen, N. C. Harris, S. Skirlo, M. Prabhu, T. Baehr-Jones, M. Hochberg, X. Sun, S. Zhao, H. Larochelle, D. Englund, M. Soljačić, Deep learning with coherent nanophotonic circuits. *Nature Photon*. **11**, 441–446 (2017).

11. A. N. Tait, T. F. de Lima, E. Zhou, A. X. Wu, M. A. Nahmias, B. J. Shastri, P. R. Prucnal, Neuromorphic photonic networks using silicon photonic weight banks. *Scientific Reports*. **7**, 1–10 (2017).

12. N. Mohammadi Estakhri, B. Edwards, N. Engheta, Inverse-designed metastructures that solve equations. *Science*. **363**, 1333–1338 (2019).

13. J. Wu, P. Cao, X. Hu, X. Jiang, T. Pan, Y. Yang, C. Qiu, C. Tremblay, Y. Su, Compact tunable silicon photonic differential-equation solver for general linear time-invariant systems. *Opt. Express, OE*. **22**, 26254–26264 (2014).

14. S. Tan, L. Xiang, J. Zou, Q. Zhang, Z. Wu, Y. Yu, J. Dong, X. Zhang, High-order all-optical differential equation solver based on microring resonators. *Opt. Lett., OL*. **38**, 3735–3738 (2013).

15. J. Lu, J. Vučković, Nanophotonic computational design. *Opt. Express*. **21**, 13351 (2013).

16. A. Y. Piggott, J. Lu, K. G. Lagoudakis, J. Petykiewicz, T. M. Babinec, J. Vučković, Inverse design and demonstration of a compact and broadband on-chip wavelength demultiplexer. *Nature Photon*. **9**, 374–377 (2015).

17. S. Molesky, Z. Lin, A. Y. Piggott, W. Jin, J. Vucković, A. W. Rodriguez, Inverse design in nanophotonics. *Nature Photon*. **12**, 659–670 (2018).

18. H. Chen, P. Verheyen, P. D. Heyn, G. Lepage, J. D. Coster, P. Absil, G. Roelkens, J. V. Campenhout, in *2016 Optical Fiber Communications Conference and Exhibition (OFC)* (2016), pp. 1–3.

19. L. Chen, P. Dong, M. Lipson, High performance germanium photodetectors integrated on submicron silicon waveguides by low temperature wafer bonding. *Opt. Express*. **16**, 11513 (2008).

20. High-performance Ge-on-Si photodetectors | Nature Photonics, (available at https://www.nature.com/articles/nphoton.2010.157).

21. X. Gan, R.-J. Shiue, Y. Gao, I. Meric, T. F. Heinz, K. Shepard, J. Hone, S. Assefa, D. Englund, Chip-integrated ultrafast graphene photodetector with high responsivity. *Nature Photonics*. **7**, 883 (2013).

22. I. Goykhman, U. Sassi, B. Desiatov, N. Mazurski, S. Milana, D. de Fazio, A. Eiden, J. Khurgin, J. Shappir, U. Levy, A. C. Ferrari, On-Chip Integrated, Silicon–Graphene Plasmonic Schottky Photodetector with High Responsivity and Avalanche Photogain. *Nano Letters*. **16**, 3005–3013 (2016).

23. A. Pospischil, M. Humer, M. M. Furchi, D. Bachmann, R. Guider, T. Fromherz, T. Mueller, CMOS-compatible graphene photodetector covering all optical communication bands. *Nature Photonics*. **7**, 892 (2013).

24. Z. Ma, K. Kikunaga, H. Wang, S. Sun, R. Amin, R. MAITI, M. H. Tahersima, H. Dalir, M. Miscuglio, V. J. Sorger, Compact Graphene Plasmonic Slot Photodetector on Silicon-on-insulator with High Responsivity. *ACS Photonics* (2020), doi:10.1021/acsphotonics.9b01452.

25. T. W. Hughes, I. A. D. Williamson, M. Minkov, S. Fan, Wave physics as an analog recurrent neural network. *Science Advances*. **5**, eaay6946 (2019).

26. V. K. Narayana, S. Sun, A.-H. A. Badawy, V. J. Sorger, T. El-Ghazawi, MorphoNoC: Exploring the design space of a configurable hybrid NoC using nanophotonics. *Microprocessors and Microsystems*. **50**, 113–126 (2017).

27. D. A. B. Miller, Attojoule Optoelectronics for Low-Energy Information Processing and Communications. *J. Lightwave Technol., JLT*. **35**, 346–396 (2017).

28. R. Amin, R. Maiti, C. Carfano, Z. Ma, M. H. Tahersima, Y. Lilach, D. Ratnayake, H. Dalir, V. J. Sorger, 0.52 V mm ITO-based Mach-Zehnder modulator in silicon photonics. *APL Photonics*. **3**, 126104 (2018).

29. R. Amin, Z. Ma, R. Maiti, S. Khan, J. B. Khurgin, H. Dalir, V. J. Sorger, Attojoule-efficient graphene optical modulators. *Appl. Opt.* **57**, D130 (2018).

30. V. J. Sorger, R. Amin, J. B. Khurgin, Z. Ma, H. Dalir, S. Khan, Scaling vectors of attoJoule per bit modulators. *J. Opt.* **20**, 014012 (2018).

31. H. K. Tsang, Y. Liu, Nonlinear optical properties of silicon waveguides. *Semicond. Sci. Technol.* **23**, 064007 (2008).

32. S. K. Pickus, S. Khan, C. Ye, Z. Li, V. J. Sorger "Silicon Plasmon Modulators: Breaking Photonic Limits" **IEEE Photonics Society**, 27, 6 (2013).

33. R. Amin, C. Suer, Z. Ma, J. Khurgin, R. Agarwal, V. J. Sorger "A Deterministic Guide for Material and Mode Dependence of On-Chip Electro-Optic Modulator Performance" **Solid-State Electronics** 136, 92-101 (2017).

34. R. Amin, J. B. Khurgin, V. J. Sorger "Waveguide-based electro-absorption modulator performance: comparative analysis" **Optics Express** 26, 11, 15445-15470 (2018)**.**

35. G. B. Folland, *Introduction to partial differential equations* (Princeton University Press, Princeton, N.J, 2nd ed., 1995).

36. S. Loranger, M. Gagné, R. Kashyap, Capacitors go optical: wavelength independent broadband mode cavity. *Opt Express*. **22**, 14253–14262 (2014).

37. A. Majumdar, A. Rundquist, Cavity-enabled self-electro-optic bistability in silicon photonics. *Opt. Lett., OL*. **39**, 3864–3867 (2014).

38. S. Sun, R. Zhang, J. Peng, V. K. Narayana, H. Dalir, T. El-Ghazawi, V. J. Sorger, MO detector (MOD): a dual-function optical modulator-detector for on-chip communication. *Opt. Express*. **26**, 8252 (2018).